# SPH modelling of hydrodynamic lubrication: laminar fluid flow-structure interaction with no-slip conditions for slider bearings

Marco Paggi[1,*], Andrea Amicarelli[2,1], Pietro Lenarda[,1]

[1]IMT School for Advanced Studies Lucca, Piazza San Francesco, 19, 55100 Lucca, Italy
[2]Ricerca sul Sistema Energetico - RSE SpA, Department SFE, via Rubattino, 54, 20134 Milan, Italy

[*]Corresponding author
marco.paggi@imtlucca.it
andrea.amicarelli@rse-web.it
pietro.lenarda@imtlucca.it

**Abstract**
The FOSS CFD-SPH code SPHERA v.9.0.0 (RSE SpA) is empowered to deal with "fluid - solid body" interactions under no-slip conditions and laminar regimes for the simulation of hydrodynamic lubrication. The code is herein validated in relation to a uniform slider bearing (i.e., for a constant lubricant film depth) and a linear slider bearing (i.e., for a film depth with a linear profile variation along the main flow direction). Validations refer to comparisons with analytical solutions, herein generalized to consider any Dirichlet boundary condition. Further, this study allows a first code validation of the "fluid-fixed frontier" interactions under no-slip conditions. With respect to the most state-of-the-art models (2D codes based on Reynolds' equation for fluid films), the following distinctive features are highlighted: (*i*) 3D formulation on all the terms of the Navier-Stokes equations for incompressible fluids with uniform viscosity; (*ii*) validations on both local and global quantities (pressure and velocity profiles; load-bearing capacity); (*iii*) possibility to simulate any 3D topology. This study also shows the advantages of using a CFD-SPH code in simulating the inertia and 3D effects close to the slider edges, and it opens new research directions overcoming the limitations of the codes for hydrodynamic lubrication based on the Reynolds' equation for fluid films. This study finally allows SPHERA to deal with hydrodynamic lubrication and empowers the code for other relevant application fields involving fluid-structure interactions (e.g., transport of solid bodies by floods and earth landslides; rock landslides). SPHERA is developed and distributed on a GitHub public repository.

# 1. Introduction

In tribology, the science and technology of interacting solid surfaces under relative motion, three lubrication regimes are defined: the full-film lubrication regime (i.e. the solid surfaces are completely separated by the fluid), the boundary lubrication regime (i.e. the solid surfaces are in direct contact by means of their asperities and possible additives) and the mixed lubrication regime (i.e. an intermediate regime). In particular, under the full-film lubrication regime, hydrodynamic lubrication is a simplification of the elasto-hydrodynamic lubrication, in case elasticity effects are negligible. Slider bearings and roller bearings are typical examples of applications related to hydrodynamic lubrication and elasto-hydrodynamic lubrication, respectively. Bearings are largely employed in several domains: electric machines, turbomachinery, internal combustion engines, electric vehicles, hydraulic systems, medicine, automation, etc. Hereafter follows a brief introduction to numerical modelling of bearings for hydrodynamic lubrication.

Williams & Symmons [1] developed a 1D CFD (Computational Fluid Dynamics) - FD (Finite Difference) model to numerically reproduce the pressure longitudinal profiles within the thin fluid film of a linear slider (or linear sliding bearing).Dobrica & Fillon [2] developed a 2D CFD-FVM (Finite Volume Method) code, alternatively using Navier-Stokes equations and Reynolds' equation for fluid films, and validated it in relation to the Rayleigh step bearings. They also highlighted the importance of modelling the advective/inertia terms, neglected by Reynolds' equation for fluid films. Similarly, Vakilian et al. [3] found that the assumption of no advection/inertia terms is responsible for up-stream under-estimations (around the leading edge) and down-stream over-predictions (around the trailing edge) in the pressure field for step bearings.

Almqvist et al. [4] provided inter-comparisons between a FD (Finite Difference) model based on Reynolds' equation for fluid films and a commercial CFD-FVM code based on Navier-Stokes equations.In a series of follow-up articles, Almqvist et al. [5,6] reported analytical solutions on pressure longitudinal profiles, velocity vertical profiles, friction force and load-bearing capacity (the frictional coefficient is the ratio between these two forces) for both a linear slider and the Rayleigh step slider, with null Dirichlet boundary conditions for pressure. They also derived the optimum geometric configuration for a linear slider to maximize the load-bearing capacity ($k$=1.189, where $k$ is the distance increase of the surface gap per unit length). Further, they analysed the effects of surface roughness by means of a homogenization technique.

Rahmani et al. [7] presented an analytical approach based on Reynolds' equation for asymmetric partially textured slider bearings with surface discontinuities, to optimize the choice of the textures parameters with respect to the load-bearing capacity and the friction force. Papadopoulos et al. [8] use a 2D CFD-FVM code to optimize micro-thrust bearings with surface texturing by means of numerical inter-comparisons.Fouflias et al. [9] used a commercial CFD-FVM code to simulate bearings with pockets/dimples and surface texturing, providing model inter-comparisons on steady-loads for different designs.

Regarding modelling of complex surface topologies, Paggi and He [10] investigated the effect of roughness on the evolution of the channel network influencing the fluid flow in the mixed lubrication regime.

Gropper et al. [11] proposed a detailed review on hydrodynamic lubrication of textured surfaces, included (multi-scale) roughness effects and cavitation. Hajishaflee et al. [12] adopted a 2D CFD-FVM model to reproduce elasto-hydrodynamic lubrication problems for rolling element bearings, including cavitation effects. Snyder and Braun [13] provided analytical solutions and 2D CFD - FVM (Finite Volume Method) results to quantify the squeezing effects on a sliding bearing load in terms of dynamic coefficients (dimensionless stiffness and damping). They also used a PRE (Perturbed Reynolds Equation) - FD model, which is based on the representation of perturbed quantities (film thickness and pressure) within Reynolds' equation, and provided three separated differential equations for static pressure, dynamic pressure associated with stiffness, and dynamic pressure associated with damping.

With respect to the state-of-the-art on CFD modelling for bearings, mostly based on 2D codes or Reynolds' simplified equation, the present study adopts a 3D CFD discretization of all the terms of the Navier-Stokes equations for incompressible fluids with uniform viscosity. It also provides validations on local quantities (pressure and velocity profiles) and it is able to take into account any 3D surface in input. In particular, the present study empowers, validates and applies the FOSS (Free/Libre and Open Source Software) CFD-SPH (Smoothed Particle Hydrodynamics) code SPHERA [14] to "fluid - solid body" interactions under no-slip conditions, simulating uniform and linear sliders. Validations are provided by comparisons with analytical solutions, here generalized to deal with non-null Dirichlet boundary conditions for pressure. Validations refer to pressure longitudinal profiles, velocity vertical profiles and load-bearing capacity. Furthermore, a demonstrative test case is simulated to represent a linear slider moving over a complex 3D surface, to show its applicability to any 3D surface input data. Beyond the new numerical developments, other code features are herein first validated, in particular the "fluid - fixed frontier" interactions under no-slip conditions. This study represents one the first applications of the SPH method to bearings. The basic features of this numerical method are briefly recalled hereafter.

Smoothed Particle Hydrodynamics (SPH) is a mesh-less CFD method, whose computational nodes are represented by numerical fluid particles. In the continuum, the functions and derivatives in the fluid dynamics balance equations are approximated by convolution integrals, which are weighted by interpolating (or smoothing functions), called kernel functions.

The integral SPH approximation ($<>_I$) of a generic function ($f$) is defined as:

$$\langle f \rangle_{I, \underline{x}_0} = \int_{V_h} f W dx^3 \tag{1}$$

where $W$ (m$^{-3}$) is the kernel function [15], $\underline{x}_0$ (m) the position of a generic computational point and $V_h$ (m$^3$) the integration volume, which is called kernel support. This is represented by a sphere of radius $2h_{SPH}$ (m), possibly truncated by the frontiers of the fluid domain.

Any first derivative of a generic function, calculated along $i$-axis, can be computed as in (1), after replacing $f$ with the targeted derivative. After integration by parts, one obtains:

$$\left\langle \frac{\partial f}{\partial x_i} \right\rangle\bigg|_{I,\underline{x}_0} = \int_{A_h} fWn_i dx^2 - \int_{V_h} f \frac{\partial W}{\partial x_i} dx^3 \qquad (2)$$

The integration also involves the surface $A_h$ (m$^2$) of the kernel support. The associated surface integral is non-zero in case of a truncated kernel support. The representation of this term noticeably differentiates SPH codes among each other [16-20].

Far from boundaries, the SPH particle approximation of (2) reads:

$$\left\langle \frac{\partial f}{\partial x_i} \right\rangle\bigg|_{\underline{x}_0} = -\sum_b f_b \frac{\partial W}{\partial x_i}\bigg|_b \omega_b \qquad (3)$$

where a summation on particle volumes $\omega$ (m$^3$) replaces the volume integral. The subscripts "$_0$" and "$_b$" refer to the computational particle and its "neighbouring particles" (fluid particles within the kernel support of the computational particle), respectively.

Usually, the approximation (3) is replaced by more complicated and accurate formulas. Further, the SPH method can also approximate a generic *n*-th derivative, analogously to (3).

Among the various numerical methods, Smoothed Particle Hydrodynamics (SPH) has several advantages: a direct estimation of free surface and phase/fluid interfaces; effective simulations of multiple moving bodies and particulate matter within fluid flows; direct estimation of Lagrangian derivatives (absence of non-linear advective terms in the balance equations); effective numerical simulations of fast transient phenomena; no meshing; simple non-iterative algorithms (in case the "Weakly Compressible" approach is adopted). On the other hand, SPH models are affected by the following drawbacks, if compared with mesh-based CFD tools: computational costs are slightly higher due to a larger stencil (around each computational particle), which causes a high number of interacting elements (neighbouring particles) at a fixed time step (nonetheless SPH codes are more suitable to parallelization); local refining of spatial resolution represents a current issue and is only addressed by few, advanced and complex SPH algorithms; accuracy is relatively low for classical CFD applications where mesh-based methods are well established (e.g., confined mono-phase flows). Detailed reviews on SPH assets and drawbacks are reported in [21-24]. Nevertheless, SPH models are effective in several, but peculiar, application fields. Some of them are here briefly recalled: flood propagation (e.g., [25, 26]); sloshing tanks (e.g., [20]); gravitational surface waves (e.g. [27]); hydraulic turbines (e.g., [28]); liquid jets (e.g., [28]); astrophysics and magneto-hydrodynamics (e.g., [29]; body dynamics in free surface flows (e.g., [30]); multi-phase and multi-fluid flows; sediment removal from water reservoirs (e.g., [31]); landslides (e.g., [32, 33]).

The paper is organized as follows. Sec. 2 reports the state-of-the-art mathematical models and analytical solutions relevant for this study. Sec. 3 derives a generalization of the analytical solutions for the linear slider to take into account non-null Dirichlet boundary conditions for pressure. Sec. 4 reports those basic features of the reference CFD-SPH code relevant for this study, whereas Sec. 5 introduces the new numerical developments of the code. Validations refer to a uniform slider (Sec. 6) and a linear slider (Sec. 7). A demonstrative test case for a linear slider with a 3D complex surface is briefly outlined in Sec. 8. The overall conclusions are synthesized in Sec. 9.

## 2. Benchmark analytical solutions

One considers a linearization of Navier-Stokes' equations for incompressible Newtonian fluids, with a fluid film flowing between two solid plates:

$$\frac{\partial u_j}{\partial x_j} = 0, \qquad \frac{\partial p}{\partial x_i} = -\frac{\partial}{\partial z}\left[\mu\left(\frac{\partial u_i}{\partial z} - \delta_{i3}\frac{\partial u_i}{\partial z}\right)\right] \qquad (4)$$

where $p$ is pressure, $\underline{u}$ is the velocity vector, $\mu$ is the dynamic viscosity, $\delta_{ij}$ is Kronecker's delta and $\underline{x}$ represents a generic position. Advective and gravity terms are herein neglected, and the viscous shear stress terms only affect the horizontal projection of momentum. Combining the above expressions, defining the fluid depth $h$ (x,y,t) and assuming the following hypotheses ($h_0$ represents the minimum value of $h$ and $L$ the upper plate length):

$$\varepsilon \equiv \frac{h_0}{L} << 1; \qquad p \propto \varepsilon^{-2}, \varepsilon \to 0 \qquad (5)$$

one obtains Reynolds' equation for fluid films:

$$\frac{\partial}{\partial x}\left(\frac{h^3}{12\mu}\frac{\partial p}{\partial x}\right)+\frac{\partial}{\partial y}\left(\frac{h^3}{12\mu}\frac{\partial p}{\partial y}\right)=\frac{(u_{s1}+u_{s2})}{2}\frac{\partial h}{\partial x}+\frac{(v_{s1}+v_{s2})}{2}\frac{\partial h}{\partial y}+\frac{\partial h}{\partial t}, \qquad 0 \leq x \leq L \qquad (6)$$

where the subscripts "$_{s1}$" and "$_{s2}$" denote the upper and lower plate, respectively. The last formula represents a 2D time-dependent equation to describe the dynamics of fluid films between two solid plates.

In case of stationary regime and uniformity along the *y*-axis, the above equation reduces to 1D Reynolds' equation for fluid films:

$$\frac{\partial}{\partial x}\left(\frac{h^3}{12\mu}\frac{\partial p}{\partial x}\right)=\frac{(u_{s1}+u_{s2})}{2}\frac{\partial h}{\partial x}, \qquad 0 \leq x \leq L \qquad (7)$$

In case of uniform viscosity, one obtains:

$$\frac{1}{6\mu}\left(h^3\frac{\partial^2 p}{\partial x^2}+3h^2\frac{\partial h}{\partial x}\frac{\partial p}{\partial x}\right)=(u_{s1}+u_{s2})\frac{\partial h}{\partial x}, \qquad 0 \leq x \leq L \qquad (8)$$

Under these assumptions, the analytical solution of any velocity profile for generic plate geometries assumes the following form:

$$u=\frac{z(z-h)}{2\mu}\frac{\partial p}{\partial x}+(u_{s1}-u_{s2})\frac{z}{h}+u_{s2} \qquad (9)$$

where velocity is proportional to the pressure derivative along *x*-axis and the mass flow rate is uniform [6]. The analytical solutions in [6] for both the uniform slider and the linear slider are reported in Sec. 2.1**Errore. L'origine riferimento non è stata trovata.** and Sec. 2.2, respectively.

**2.1 Uniform slider**

The uniform slider is featured by a uniform value for the fluid depth *h*. This geometrical configuration implies a 1D Laplace equation for pressure, which assumes a linear horizontal profile:

$$h=B \Rightarrow \frac{\partial^2 p}{\partial x^2}=0 \Rightarrow p(x)=p_0+(p_L-p_0)\frac{(x-x_0)}{(x_L-x_0)} \qquad (10)$$

where the subscripts "$_L$" and "$_0$" denote the plate edges.

Considering Eq. (9), the analytical solution for the velocity profiles within a uniform slider is:

$$u = \frac{z(z-h)}{2\mu}\frac{(p_L - p_0)}{L} + (u_{s1} - u_{s2})\frac{z}{h} + u_{s2} \tag{11}$$

The load-bearing capacity $l_c$ represents the hydrodynamic thrust exerted on the plate. On the uniform slider, the analytical solution for the load-bearing capacity (per unit width) reads:

$$l_c = \frac{p_0 + p_L}{2} L \tag{12}$$

The following non-dimensional quantities are defined for pressure, velocity magnitude and time:

$$C_p \equiv \frac{p}{\frac{1}{2}\rho(U^*)^2}, \quad U^* \equiv |u_{s,1}| + |u_{s,2}|, \quad T \equiv \frac{2tU^*}{h_0} \tag{13}$$

with $C_p$ denoting the pressure coefficient.

The non-dimensional load-bearing capacity $L_C$ is defined as follows:

$$L_C \equiv \frac{l_c}{\frac{1}{2}\rho(U^*)^2 L} \tag{14}$$

For a uniform slider, the analytical solution for $L_C$ is equal to the average of the pressure coefficient values provided as boundary conditions:

$$L_C \equiv \frac{C_{p,0} + C_{p,L}}{2} \tag{15}$$

**2.2 Linear slider**

Within a linear slider, the depth rate of change of the lubricant along the *x*-axis is constant:

$$h(x) \equiv h_0\left(1 + k - k\frac{x}{L}\right) \tag{16}$$

Almqvist [6] assumed null pressure at the edges of the linear slider and provided analytical solutions for the longitudinal profile of pressure (which is uniform along the vertical direction), the vertical profiles of the horizontal velocity, and the load-bearing capacity:

$$p(x) = \frac{6\mu(u_{s,1}+u_{s,2})L}{kh_0^2}\left(\frac{1}{\left(1+k-k\frac{x}{L}\right)} - \frac{1}{\left(1+k-k\frac{x}{L}\right)^2}\frac{(1+k)}{(2+k)} - \frac{1}{(2+k)}\right)$$

$$u(x,z) = \left\{\left[1-\frac{2h_0}{h}\frac{(1+k)}{(2+k)}\right]3(u_{s,1}+u_{s,2})\right\}\frac{z(z-h)}{h^2} + (u_{s,1}-u_{s,2})\frac{z}{h} + u_{s,2} \qquad (17)$$

$$l_c = \frac{6\mu(u_{s,1}+u_{s,2})L^2}{h_0^2}\left[\frac{1}{k^2}\ln(1+k) - \frac{2}{k(2+k)}\right]$$

## 3. Generalized analytical solutions for a linear slider

Assuming a vanishing pressure at the edges of a linear slider is a theoretical simplification, which does not take into account the over-pressure and the under-pressure zones determined by the interaction between the fluid flow and the solid plate at its leading and trailing edges, respectively. The analytical solutions of Almqvist et al. [6] are herein generalized to impose non-null Dirichlet's boundary conditions for pressure at the edges of a linear slider, to cope with more general and practical configurations. No matter about the particular value of pressure at the edges of the plate, the longitudinal pressure profile reads [6]:

$$p(x) = 6\frac{\mu UL}{kh_0 h} + \frac{L}{2kh_0}\frac{C_1}{h^2} + C_2 \qquad (18)$$

where $C_1$ and $C_2$ are integration constants.

The velocity scale $U$ is the $x$-component of the vector summation of the velocities of the solid plates:

$$U_s \equiv u_{s,1} + u_{s,2} \qquad (19)$$

One imposes generic now non-null Dirichlet's boundary conditions for pressure at the slider edges:

$$p(0) = 6\frac{\mu U_s L}{kh_0^2(1+k)} + \frac{L}{2kh_0^3(1+k)^2}C_1 + C_2 = p_0 \qquad (20a)$$

$$p(L) = 6\frac{\mu U_s L}{k h_0^2} + \frac{L}{2 k h_0^3} C_1 + C_2 = p_L \qquad (20b)$$

Solving the first formula of (20) for $C_2$ and replacing this expression in the second formula of (20), one obtains the following expression for the integration constant $C_1$:

$$C_1 = -2h_0 \frac{(1+k)}{(2+k)} \left[ 6\mu U_s + \frac{(p_0 - p_L) h_0^2}{L} (1+k) \right] \qquad (21)$$

Replacing $C_1$ into the second expression of (20), one obtains the integration constant $C_2$:

$$C_2 = -\frac{L}{h_0^2 k(k+2)} \left[ 6\mu U_s + \frac{(p_L - p_0) h_0^2}{L} (1+k)^2 \right] + p_L \qquad (22)$$

Replacing the expressions for the integration constants (21)-(22) and the fluid depth (16) into Eq. (18), the pressure profile assumes the following form:

$$p(x) = \frac{6\mu(u_{s,1} + u_{s,2})L}{k h_0 h} + \frac{L}{2 k h_0 h^2} \left( -2h_0 \frac{(1+k)}{(2+k)} \left[ 6\mu(u_{s,1} + u_{s,2}) + \frac{(p_0 - p_L) h_0^2}{L} (1+k) \right] \right) +$$

$$- \frac{L}{k h^2(k+2)} \left[ 6\mu(u_{s,1} + u_{s,2}) + \frac{(p_L - p_0) h_0^2}{L} (1+k)^2 \right] + p_L \qquad (23)$$

After minor arrangements, one obtains the following analytical solution for the longitudinal profile of pressure within a linear slider under generic non-null Dirichlet boundary conditions:

$$p(x) = \frac{6\mu(u_{s,1} + u_{s,2})L}{k h_0^2} \left( \frac{1}{\left(1 + k - k\frac{x}{L}\right)} - \frac{1}{\left(1 + k - k\frac{x}{L}\right)^2} \frac{(1+k)}{(2+k)} - \frac{1}{(2+k)} \right) +$$

$$- (p_0 - p_L) \frac{(1+k)^2}{k(2+k)} \left[ \frac{1}{\left(1 + k - k\frac{x}{L}\right)^2} - 1 \right] + p_L \qquad (24)$$

It is remarkable to note that the first term of the Right Hand Side (RHS) of Eq. (24) represents the solution in the theoretical case of null pressure at the edges of the linear slider, Eq. (17). This is directly proportional to the viscosity, the velocity scale, the slider length, and it is inversely proportional to the square of the minimum fluid depth. The expression within brackets only depends upon the geometric parameter $k$ and the normalized distance from a slider edge, $x/L$. The second term on the RHS of Eq. (24) is directly proportional to the difference between the edge pressure values and depends on $k$ and $x/L$. The third term on the RHS of Eq. (24) is represented by the pressure value at the slider edge, the most distant one from the origin of the reference system.

Under normal conditions (pressure at the leading edge "$0$" is higher than pressure at the trailing edge "$L$"), the second and third terms of the RHS of Eq. (24) raise pressure levels. The second term also shifts the pressure peak towards the body edge with higher pressure.

The integration over the plate length of the analytical solution for pressure (Eq. (24)) provides the expression for the load-bearing capacity:

$$l_c = \int_0^L p(x)dx = \frac{6\mu(u_{s,1}+u_{s,2})L^2}{h_0^2}\left[\frac{1}{k^2}\ln(1+k) - \frac{2}{k(2+k)}\right] +$$

$$-(p_0 - p_L)\frac{(1+k)^2}{k(2+k)}\int_0^L \left[\frac{1}{\left(1+k-k\frac{x}{L}\right)^2} - 1\right]dx + p_L L \quad (25)$$

The second to last term in Eq. (25) assumes the following expression:

$$-(p_0 - p_L)\frac{(1+k)^2}{k(2+k)}\left[\int_0^L \left(1+k-k\frac{x}{L}\right)^{-2}dx - \int_0^L dx\right] =$$

$$= (p_0 - p_L)L\frac{(1+k)^2}{k(2+k)}\frac{k^2}{k(1+k)} = (p_0 - p_L)L\frac{(1+k)}{(2+k)} \quad (26)$$

And, once it is introduced back into Eq. (25), the load-bearing capacity reads:

$$l_c = \frac{6\mu(u_{s,1}+u_{s,2})L^2}{h_0^2}\left[\frac{1}{k^2}\ln(1+k) - \frac{2}{k(2+k)}\right] + (p_0 - p_L)L\frac{(1+k)}{(2+k)} + p_L L \quad (27)$$

and its non-dimensional formulation renders:

$$L_C = 2\left\{\frac{6\upsilon(u_{s,1}+u_{s,2})L}{(|u_{s,1}|+|u_{s,2}|)^2 h_0^2}\left[\frac{1}{k^2}\ln(1+k) - \frac{2}{k(2+k)}\right] + \frac{(p_0-p_L)}{\rho(|u_{s,1}|+|u_{s,2}|)^2}\frac{(1+k)}{(2+k)} + \frac{p_L}{\rho(|u_{s,1}|+|u_{s,2}|)^2}\right\} \quad (28)$$

A generic analytical form for the 2D field of the *x*-component of velocity ($u$) is expressed by Almqvist et al. [14]:

$$u(x,z) = \frac{z(z-h)}{2\mu}\left(\frac{6\mu U_s}{h^2} + \frac{C_1}{h^3}\right) + \frac{z}{h}(u_1 - u_2) + u_2 \qquad (29)$$

Considering the integration constant $C_1$ as provided by Eq. (21), the generalized analytical solution for the 2D velocity field assumes the following expression:

$$u(x,z) = \left\{\left[1 - \frac{2h_0}{h}\frac{(1+k)}{(2+k)}\right]3U_s - \frac{(1+k)^2}{(2+k)}\frac{(p_0 - p_L)h_0^3}{L\mu h}\right\}\frac{z(z-h)}{h^2} + (u_1 - u_2)\frac{z}{h} + u_2 \qquad (30)$$

The configuration of the linear slider used to validate the code SPHERA (Sec. 6.1) is featured by $h$ dependent both on $x$ and t (Figure 1, right panel). This configuration needs to be related to the configuration associated with the above analytical solution (Figure 1, left panel), by comparing to an intermediate configuration (Figure 1, centre panel) and considering the variable changes for the horizontal components of position, velocity and pressure gradient, and for the fluid depth:

$$x^* = x - u_{s,1}^* t; \qquad x^{**} + u_{s,1}^* t = L - x^* + u_{s,1}^* t \Rightarrow x^{**} = L - x^* - 2u_{s,1}^* t = L - x - u_{s,1}^{**} t$$

$$u^*\left(x^* + u_{s,1}^* t\right) = u(x) - u_{s,2}; \quad u^{**}\left(x^{**} + u_{s,1}^{**} t\right) = -u^*\left(x^* + u_{s,1}^* t\right) = -u(x) + u_{s,2}$$

$$\left.\frac{\partial p^{**}}{\partial x^{**}}\right|_{(x^{**} + u_{s,1}^{**} t)} = -\left.\frac{\partial p^*}{\partial x^*}\right|_{(x^* + u_{s,1}^* t)} = -\left.\frac{\partial p}{\partial x}\right|_x \qquad (31)$$

$$h = h_0\left[1 + k\left(1 - \frac{x}{L}\right)\right] = h_0\left[1 + k\left(1 - \frac{x^* + u_{s,1}^* t}{L}\right)\right] = h_0\left[\left(1 + k\left(\frac{x^{**} - u_{s,1}^{**} t}{L}\right)\right)\right]$$

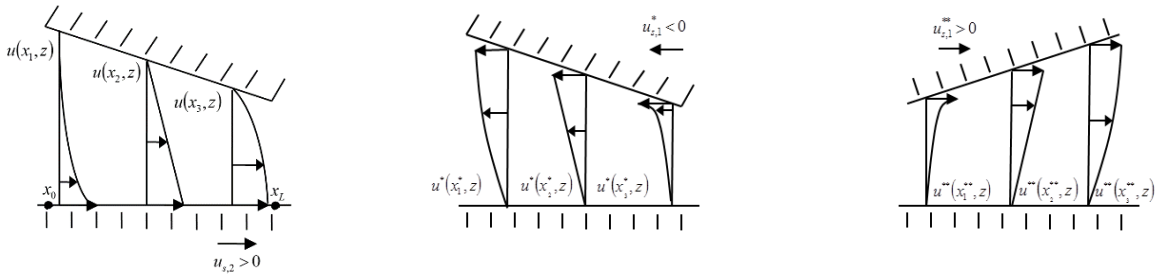

Figure 1. Linear slider. Left panel: basic configuration associated with the analytical solutions. Centre panel: intermediate configuration ("*"). Right panel: configuration associated with the code validation ("**").

# 4. Description of the SPH computational framework

The reference code SPHERA used in this study [14] has been applied to floods (with transport of solid bodies, bed-load transport and a domain spatial coverage up to some hundredths of squared kilometres), fast landslides and wave motion, sediment removal from water reservoirs, sloshing tanks. SPHERA is featured by: a scheme for dense granular flows [34]; a scheme for the transport of solid bodies in free surface flows [30]; a scheme for a boundary treatment ("DB-SPH") based on discrete surface and volume elements, and on a 1D Linearized Partial Riemann Solver coupled with a MUSCL (Monotonic Upstream-Centered Scheme for Conservation Laws) spatial reconstruction scheme [20]; a scheme for a 2D erosion criterion [31]; a scheme for a boundary treatment ("semi-analytic approach or SA-SPH" for simplicity of notation) based on volume integrals, numerically computed outside of the fluid domain [35].

This section only reports the basic features of the code relevant for this study: the balance equations for fluid (Sec.4.1) and body (Sec.4.2) dynamics, the semi-analytic approach for treating fixed boundaries (Sec.4.1) and the 2-way interaction terms related to the fluid-body interactions (Sec.4.3).

## 4.1 SPH approximation of the balance equations for fluid dynamics and the boundary treatment scheme called "semi-analytic approach"

The numerical scheme for the main flow is a Weakly-Compressible (WC) SPH model, which takes benefit from a boundary treatment for fixed boundaries based on the semi-analytic approach of Vila [36], as developed by Di Monaco et al. [35].

One considers Navier-Stokes' momentum and continuity equations for incompressible fluids with uniform viscosity:

$$\frac{du_i}{dt} = -\frac{1}{\rho}\frac{\partial p}{\partial x_i} - \delta_{i3}g + \nu\frac{\partial^2 u_i}{\partial x_j^2}, \qquad i=1,2,3$$

$$\frac{d\rho}{dt} = -\rho\nabla \cdot \underline{u}$$

(32)

One needs to compute Eq. (32) at each fluid particle position by using the SPH formalism and taking into account the boundary terms (fluid-frontier and fluid-body interactions), as described in the following.

One considers the discretization of Eq. (32), as provided by the SPH approximation of the first derivative of a generic function (*f*), according to the semi-analytic approach [36] ("$_{SA}$"):

$$\left\langle \frac{\partial f}{\partial x_i} \right\rangle_{SA,0} = -\sum_b (f_b - f_0) \frac{\partial W_b}{\partial x_i} \omega_b - \int_{V_h'} (f - f_0) \frac{\partial W}{\partial x_i} dx^3 \qquad (33)$$

The inner fluid domain here involved is filled with numerical particles. At boundaries, the kernel support is (formally) not truncated because it can partially lie outside the fluid domain. In other words, the summation in Eq. (33) is performed over all the fluid particles "$_b$" (neighbouring particles with volume $\omega$) in the kernel support of the computational fluid particle ("$_0$"). At the same time, the volume integral in Eq. (33) represents the boundary term, which is a convolution integral on the truncated portion of the kernel support. In this fictitious and outer volume $V_h'$ (m$^3$), one needs to define the generic function *f* (pressure, velocity or density alternatively).

The semi-analytic approach ("$_{SA}$") (as developed in [35]) introduces the following linearization and assumptions to compute *f* in $V_h'$:

$$\left\langle \frac{\partial f}{\partial x_i} \right\rangle_{SA} = -\sum_b (f_b - f_0) \frac{\partial W_b}{\partial x_i} \omega_b - \int_{V_h'} f_{SA} \frac{\partial W}{\partial x_i} dx^3 - \int_{V_h'} \left.\frac{\partial f}{\partial x_i}\right|_{SA} (\underline{x} - \underline{x}_0) \frac{\partial W}{\partial x_i} dx^3 \qquad (34)$$

The peculiar "$_{SA}$" values of the functions and derivatives within $V_h'$ are assigned to represent a null normal gradient of reduced pressure at the frontier interface (while considering uniform density):

$$p_{SA} = p_0, \qquad \left\langle \frac{\partial p}{\partial x_i} \right\rangle_{SA} = -\delta_{i3} g; \qquad \rho_{SA} = \rho_0, \qquad \left\langle \frac{\partial \rho}{\partial x_i} \right\rangle_{SA} = 0 \qquad (35)$$

At the same time, the model sets free-slip conditions when estimating velocity at boundaries. The velocity vector is taken as uniform in the outer part of the kernel support. Here $\underline{u}_{SA}$ is decomposed in the sum of a vector normal to boundary $(\underline{u}_{SA,n})$ and a tangential vector $(\underline{u}_{SA,T})$. The first is represented as a linear extrapolation from the computational fluid particle velocity. The latter is equal to its analogous vector of the same computational fluid particle (the subscript "$w$" refers to a generic frontier), in case of no-slip conditions:

$$\underline{u}_{SA,T} = \underline{u}_{w,T} - (\underline{u}_{0,T} - \underline{u}_{w,T}) = 2\underline{u}_{w,T} - \underline{u}_{0,T} = [(2\underline{u}_w - \underline{u}_0) \cdot \underline{t}_w]\underline{t}_w$$
$$\underline{u}_{SA,n} = \underline{u}_{w,n} - (\underline{u}_{0,n} - \underline{u}_{w,n}) = 2\underline{u}_{w,n} - \underline{u}_{0,n} = [(2\underline{u}_w - \underline{u}_0) \cdot \underline{n}_w]\underline{n}_w \quad (36)$$

The velocity field within the truncated portion of the kernel support assumes is defined as follows:

$$\underline{u}_{SA} = \underline{u}_{SA,n} + \underline{u}_{SA,T} = [(2\underline{u}_w - \underline{u}_0) \cdot \hat{\underline{n}}_w]\hat{\underline{n}}_w + [(2\underline{u}_w - \underline{u}_0) \cdot \underline{t}_w]\underline{t}_w = 2\underline{u}_w - \underline{u}_0$$

$$\left\langle \frac{\partial u_i}{\partial x_i} \right\rangle_{SA} = 0 \quad (37)$$

where $\underline{n}$ is the normal vector of the wall surface, according to its local orientation.

At this point, one can write the continuity equation for a Weakly Compressible SPH model (Einstein's notation works for the subscript "$j$"), using the semi-analytic approach for the boundary integral term (second term on the Right Hand Side):

$$\left\langle \frac{d\rho}{dt} \right\rangle_0 = \sum_b \rho_b (u_{b,j} - u_{0,j}) \frac{\partial W}{\partial x_j}\bigg|_b \omega_b + 2\rho_0 \int_{V_h'} [(\underline{u}_w - \underline{u}_0) \cdot \underline{n}] n_j \frac{\partial W}{\partial x_j} dx^3 + C_s \quad (38)$$

where $C_s$ (kg×m$^{-3}$×s$^{-1}$) represents a "fluid-body" interaction term.

On the other hand, one can analogously derive the approximation of the momentum equation (the notation "$\langle \ \rangle$" indicates the SPH particle -discrete- approximation):

$$\left\langle \frac{du_i}{dt} \right\rangle_0 = -\delta_{i3} g + \sum_b \left( \frac{p_b}{\rho_b^2} + \frac{p_0}{\rho_0^2} \right) \frac{\partial W}{\partial x_i}\bigg|_b m_b + 2\frac{p_0}{\rho_0} \int_{V_h'} \frac{\partial W}{\partial x_i} dx^3 - \upsilon_M \sum_b \frac{m_b}{\rho_0 r_{0b}^2} (u_b - u_0) \cdot (\underline{x}_b - \underline{x}_0) \frac{\partial W}{\partial x_i}\bigg|_b +$$
$$- 2\upsilon_M (\underline{u}_w - \underline{u}_0) \cdot \int_{V_h'} \frac{1}{r_{0w}^2} (\underline{x} - \underline{x}_0) \frac{\partial W}{\partial x_i}\bigg|_b dx^3 + \underline{a}_s - 4\sum_b \left( \frac{m_b \nu_0 \nu_b}{\nu_0 \rho_0 + \nu_b \rho_b + \varepsilon_{\nu\rho}} u_{i,0b} \frac{|\nabla W|_b}{r_{0,b}(r_{0,b}^2 + \varepsilon_{r,2})} r_{0,b}^2 \right) + \quad (39)$$
$$+ 2\nu_0 (\underline{u}_w - \underline{u}_0) \int_{V_h'} \frac{1}{r} \left|\frac{\partial W}{\partial r}\right|_b dx^3$$

where $\underline{a}_s$ (m×s$^{-2}$) is a the acceleration term due to the fluid-body interactions, $\nu_M$ (m$^2$×s$^{-1}$) is the artificial viscosity [15], $m$ (kg) the particle mass and $r$ (m) the relative distance between the neighbouring and the computational particle.

The last two terms of the RHS of Eq. (39) represent the inner and boundary components of the shear stress gradient term. They are presented in [35] but had never been validated on a test case where their role was non-negligible. This study allowed fixing several bugs in the 3D implementation of these terms which could not be used with the release SPHERA v.8.0 [14].

Finally, a barotropic equation of state (EOS) is linearized as follows:

$$p \cong c_{ref}^2 (\rho - \rho_{ref}) \tag{40}$$

The artificial sound speed $c$ (m/s, "$_{ref}$" stands for a reference state) is usually 10 times higher than the maximum fluid velocity (WC approach) to obtain a density relative error within 1%. However, Monaghan [15] demonstrated this statement by assuming that the maximum pressure coefficient is 2. Each test case of this study considers an external force continuously applied to the fluid so that the maximum value of $C_p$ is noticeably higher than 2. A generalization for the definition of the sound speed value is here obtained assuming that $c$ is $5C_{p,max}$ times higher than the maximum fluid velocity.

The reader is referred to [30] and in [35] for more details.

**4.2 SPH balance equations for rigid body transport**

Body dynamics is ruled by Euler-Newton equations, whose discretization takes advantage from the SPH formalism and the coupling terms derived in the following sections:

$$\frac{d\underline{u}_{CM}}{dt} = \frac{\underline{F}_{TOT}}{m_B}, \qquad \frac{d\underline{x}_{CM}}{dt} = \underline{u}_{CM}$$

$$\frac{d\underline{\chi}_B}{dt} = \underline{\underline{I}}_C^{-1} \left[ \underline{M}_{TOT} - \underline{\chi}_B \times (\underline{\underline{I}}_C \underline{\chi}_B) \right], \qquad \frac{d\underline{\alpha}}{dt} = \underline{\chi}_B \tag{41}$$

Here the subscript "$_B$" refers to a generic computational body and "$_{CM}$" to its centre of mass. The first two formulae of Eq. (41) represent the balance equations for the momentum and the time law for the position of the body barycentre. $\underline{F}_{TOT}$ (kg×m×s$^{-2}$) is the global/resultant force exerted on the solid. The last two formulae of Eq. (41) express the balance equation of the angular momentum

-$\chi_B$ (rad×s$^{-1}$) denotes the angular velocity of the generic body- and the time evolution of the solid orientation -$\alpha$ (rad) is the vector of Euler angles lying between the body axes and the global reference system axes-. $\underline{M}_{TOT}$ (kg×m$^2$×s$^{-2}$) represents the associated torque acting on the body and $\underline{\underline{I_C}}$ (kg×m$^2$) is the matrix of the moments of inertia of the computational body (Einstein's notation applies to the subscript "$l$"):

$$I_{c,ij} = \int_{V_B} \rho(r_l^2 \delta_{ij} - r_i r_j) dV = \begin{cases} \int_{V_B} \rho(r_k^2 + r_n^2) dV, i = j; k, n \neq i \\ -\int_{V_B} \rho(r_i r_j) dV, i \neq j \end{cases} \quad (42)$$

In this sub-section, $\underline{r}$ implicitly represents the relative distance from the body centre of mass.

In order to solve the system (41), we need to model the global force and torque, as described in the following.

The resultant force is composed of several terms:

$$\underline{F}_{TOT} = \underline{G} + \underline{P}_F + \underline{T}_F + \underline{P}_S + \underline{T}_S, \qquad \underline{T}_F + \underline{T}_S \cong 0 \quad (43)$$

where $\underline{G}$ (kg×m×s$^{-2}$) represents the gravity force, whereas $\underline{P}_F$ (kg×m×s$^{-2}$) and $\underline{T}_F$ (kg×m×s$^{-2}$) the vector sums of the pressure and shear forces provided by the fluid. Analogously, $\underline{P}_S$ (kg×m×s$^{-2}$) and $\underline{T}_S$ (kg×m×s$^{-2}$) are the vector sums of the normal and the shear forces provided by other bodies or boundaries (solid-solid interactions). In case of inertial and quasi-inertial fluid flows, turbulence schemes and tangential stresses are not mandatory (simplifying hypothesis).

The fluid-solid interaction is expressed by the hydrodynamic thrust:

$$\underline{P}_F = \sum_s p_s A_s \underline{n}_s \quad (44)$$

The computational body is numerically represented by solid volume elements, here called (solid) "body particles" ("$s$"). Some of them describe the body surface and are referred to as "surface body particles". These particular elements are also characterized by an area and a vector $\underline{n}$ of norm 1, is perpendicular to the body face of the particle (it belongs to) and pointing outward the fluid domain (inward the solid body). The pressure of a body particle is computed as described in Sec.4.3.

The torque in (41) is discretized as the summation of each vector product between the relative position $\underline{r}_s$, of a surface body particle with respect to the body centre of mass, and the corresponding total particle force:

$$\underline{M}_{TOT} = \sum_s \underline{r}_s \times \underline{F}_s \qquad (45)$$

Time integration of (41) is performed using a Leapfrog scheme synchronized with the fluid dynamics balance equations. This means that the body particle pressure is computed simultaneously to the fluid pressure, so that this parameter is staggered of around $dt/2$ with respect to all the other body particle parameters.

After time integration, the model provides the velocity of a body particle as the vector sum of the velocity of the corresponding body barycentre and the relative velocity:

$$\underline{u}_s = \underline{u}_{CM} + \underline{\chi}_B \times \underline{r}_s \qquad (46)$$

Finally, the model updates the body particle normal vectors and absolute positions, according to the following kinematics formulas - $\underline{d\alpha}$ (rad) is the vector increment in Euler's angles during the current time step and $R_{ij}$ is the body rotation matrix-:

$$\underline{n}_s(t+dt) = \underline{\underline{R}}_B \underline{n}_s(t), \qquad \underline{x}_s(t+dt) = \underline{x}_{CM}(t+dt) + \underline{\underline{R}}_B \underline{r}_s(t)$$

$$\underline{\underline{R}}_B = \underline{\underline{R}}_x \underline{\underline{R}}_y \underline{\underline{R}}_z, \qquad \underline{d\alpha}_B = \underline{\chi}_B dt$$

$$\underline{\underline{R}}_x = \begin{bmatrix} 1 & 0 & 0 \\ 0 & \cos(d\alpha_x) & -\sin(d\alpha_x) \\ 0 & \sin(d\alpha_x) & \cos(d\alpha_x) \end{bmatrix}, \quad \underline{\underline{R}}_y = \begin{bmatrix} \cos(d\alpha_y) & 0 & \sin(d\alpha_y) \\ 0 & 1 & 0 \\ -\sin(d\alpha_y) & 0 & \cos(d\alpha_y) \end{bmatrix}, \quad \underline{\underline{R}}_z = \begin{bmatrix} \cos(d\alpha_z) & -\sin(d\alpha_z) & 0 \\ \sin(d\alpha_z) & \cos(d\alpha_z) & 0 \\ 0 & 0 & 1 \end{bmatrix}$$

(47)

More details are available in Amicarelli et al. [30].

**4.3 Fluid-body interaction terms**

The fluid-body interaction terms rely on the boundary technique introduced by Adami et al. [16], as implemented and adapted for free-slip conditions by Amicarelli et al. [30]. If the boundary is fixed, then this method can be interpreted as a discretization of the semi-analytic approach used to treat fluid-boundary interactions (Sec.4.1). The outer domain of () is herein represented by all the body particles inside the kernel support of the computational fluid particle. Further, Adami et al. [16] introduced a new term, related to the acceleration of the fluid-solid interface, which influences the estimation of body particle pressure.

The fluid-body interaction term in the momentum equation (39) assumes the form:

$$\underline{a}_s = \sum_s \left( \frac{p_s + p_0}{\rho_0^2} \right) W_s' m_s \tag{48}$$

More details are available in Amicarelli et al. [30]. The new formulation for the pressure of a generic body particle under no-slip conditions is presented in Sec.5.1.

**4.4 Time integration schemes**

Time integration is ruled by a second-order Leapfrog scheme (stability analysis and time integration schemes in SPH modelling are discussed in [37], as described in [35] and [30]:

$$x_i\big|_{t+dt} = x_i\big|_t + u_i\big|_{t+dt/2} dt, \qquad i = 1,2,3$$

$$u_i\big|_{t+dt/2} = u_i\big|_{t-dt/2} + \left\langle \frac{du_i}{dt} \right\rangle\bigg|_t dt, \qquad i = 1,2,3 \tag{49}$$

$$\rho\big|_{t+dt} = \rho\big|_t + \left\langle \frac{d\rho}{dt} \right\rangle\bigg|_{t+dt/2} dt$$

Time integration is constrained by the following stability criteria [34]:

$$dt = \min_0 \left\{ C_v \frac{2h^2}{\nu} ; CFL \frac{2h}{c + |\underline{u}|} \right\} \tag{50}$$

where *CFL* is the Courant-Friedrichs-Lewy number. Following [16], the viscous term stability parameter is set to $C_v=0.05$.

# 5. New developments for the mathematical and the numerical models: shear stress gradient term and no-slip conditions for fluid-body interactions

This section describes the new numerical developments implemented into the code SPHERA [14] by RSE SpA to describe the shear stress gradient term and no-slip conditions at the interface between the liquid domain and a mobile solid body (i.e. fluid-body interactions). The new coupling terms are introduced in the momentum (Sec. 5.1) and the continuity (Sec. 5.2) equations.

## 5.1 Fluid-body coupling terms for the momentum equation

Modelling the shear stress gradient term in the Navier-Stokes equation requires the introduction of an additional fluid-body coupling term in the RHS of the momentum equation. The discretization of this coupling term, which takes into account the shear stress exchanged at the fluid-body interface, needs to be coherent with the SPH particle approximation in the inner domain and assumes the following form:

$$2\nu_0 \sum_s \frac{(u_{s0} - u_0)}{r_{0s}} \left.\frac{\partial W}{\partial r}\right|_s \omega_s \qquad (51)$$

where the subscript "$s_0$" denotes a generic solid-fluid inter-particle interaction and the solid particle volume is computed as follows:

$$\omega_s = \frac{\omega_0}{\left(\frac{\omega_0}{\omega_s}\right)} = \frac{\left(\frac{m_0}{\rho_{ref}}\right)}{\left(\frac{dx}{dx_s}\right)^D} \qquad (52)$$

The inter-particle velocity $\underline{u}_{s0}$ represents the field of the fluid velocity virtually reconstructed within the portion of the kernel support, which is truncated by the solid body.

The component of the inter-particle velocity, which is normal to the interface, guarantees no mass penetration (symmetric conditions) at the interface:

$$\underline{u}_{s0} = \underline{u}_{s0,n} + \underline{u}_{s0,T} = [(2\underline{u}_s - \underline{u}_0) \cdot \underline{n}_s]\underline{n}_s + [(2\underline{u}_s - \underline{u}_0) \cdot \underline{t}_s]\underline{t}_s = 2\underline{u}_s - \underline{u}_0 \quad (53)$$

Under no-slip conditions, the component of the inter-particle velocity, which is tangential (subscript "$_T$") to the interface guarantees a uniform velocity gradient around the interface (if its position is assumed to be the average of the positions of the interacting particles):

$$\underline{u}_{s0,T} = \underline{u}_{s,T} - (\underline{u}_{0,T} - \underline{u}_{s,T}) = 2\underline{u}_{s,T} - \underline{u}_{0,T} = [(2\underline{u}_s - \underline{u}_0) \cdot \underline{t}_s]\underline{t}_s \quad (54)$$

where the unit vector $\underline{t}$ is tangential to the interface.

Under no-slip conditions, the difference between the inter-particle velocity and the fluid particle velocity in Eq. (51) is expressed as follows:

$$\underline{u}_{SA} - \underline{u}_0 = 2(\underline{u}_w - \underline{u}_0) \quad (55)$$

No-slip conditions also affect the formulation of the pressure gradient coupling term. The pressure value of the generic neighbouring surface body particle "$_s$" in Eq. (48) depends on the particular computational fluid particle "$_0$" we are considering, so that we can refer to the interaction subscript "$_{s,0}$".

The pressure value of the generic neighbouring surface body particle "$_s$" is derived as follows. Consider a generic point at a generic fluid-body interface. In case of free-slip conditions, the normal projection of the acceleration on the fluid side ("$_f$") and on the solid side ("$_w$") are equal (in-built motion in the direction normal to the interface):

$$\left(\frac{d\underline{u}_f}{dt}\right) \cdot \underline{n}_w = \left(-\frac{1}{\rho_f}\nabla p_f + \underline{g}\right) \cdot \underline{n}_w = \underline{a}_w \cdot \underline{n}_w \quad (56)$$

The "wall" acceleration at the position of a generic body particle can then be derived by linearizing Eq. (56). This depends on the particular computational fluid particle "$_0$" we are considering, so that we can refer to the interaction subscript "$_{s,0}$":

$$\int_0^s \left(-\frac{1}{\rho_f}\nabla p_f\right) \cdot \underline{n}_w dl_n = \int_0^s (\underline{a}_w - \underline{g}) \cdot \underline{n}_w dl_n \Rightarrow p_{s0} \approx p_0 + \rho_0[(\underline{g} - \underline{a}_s) \cdot \underline{n}_w][(\underline{x}_s - \underline{x}_0) \cdot \underline{n}_w] \quad (57)$$

where $\underline{dl}_n$ is a vectorial length element along the centreline of the two particles, projected along the wall element normal:

$$\int_0^s \left(-\frac{1}{\rho_f}\nabla p_f\right)\underline{dl}_n = \int_0^s (\underline{a}_w - \underline{g})\underline{dl}_n \Rightarrow p_{s0} \approx p_0 + \rho_0[(\underline{g}-\underline{a}_s)\cdot(\underline{x}_s - \underline{x}_0)] \quad (58)$$

One applies a SPH interpolation over all the pressure values estimated according to [16] to derive a unique pressure value for a body particle. In case of no-slip conditions, Eq. (58) assumes the following form [16]:

$$p_s = \frac{\sum_0 p_{s0} W_{s0}\left(\frac{m_0}{\rho_0}\right)}{\sum_0 W_{s0}\left(\frac{m_0}{\rho_0}\right)} = \frac{\sum_0 [p_0 + \rho_0(\underline{g}-\underline{a}_s)\cdot\underline{r}_{s0}] W_{s0}\left(\frac{m_0}{\rho_0}\right)}{\sum_0 W_{s0}\left(\frac{m_0}{\rho_0}\right)} \quad (59)$$

**5.2 Fluid-body coupling term for the continuity equation**

Modelling the shear stress gradient term in the Navier-Stokes equation requires the introduction of an additional fluid-body coupling term in the RHS of the continuity equation. The discretization of this term, which is coherent with the definition of the inter-particle velocity under no-slip conditions (Sec.5.1), assumes the following form:

$$C_s = 2\rho_0 \sum_s (\underline{u}_{s0}-\underline{u}_0)\cdot\nabla W_s \omega_s \quad (60)$$

**6 Validation: uniform slider**

Barkley and Tuckerman [38] defined the Reynolds' number ($Re$) for the configuration of a slider bearing according to the following expression:

$$\text{Re} = \frac{h(u_{s,1}+u_{s,2})}{2\nu} \quad (61)$$

As $h$ is the gap between the upper "$s,1$" and the lower "$s,2$" solid plates of the bearing, then the length scale in the Reynolds' number is half the fluid depth. According to [38], the 1D infinitive linear slider shows a laminar regime for $Re\leq290$. As the current validation refers to a finite slider, a lower value ($Re=100$) is used here to guarantee a laminar regime along the 95% of the slider length (far enough from the leading and the trailing edges of the mobile plate).

The initial fluid velocity is the average velocity of the solid plates ($u_{f,0}=u_{s,1}/2$). The reference velocity is $U^*=50$m/s. A high viscosity liquid allows representing a motor oil and makes the presence of the artificial viscosity (Sec.4.1) irrelevant.

Omitting gravity is equivalent and alternative to replacing pressure with the reduced pressure, which is defined as the difference between pressure and its hydrostatic component. Under the hypotheses of Reynolds' equation for fluid films, the film depth $h$ should tend to zero (or practically be negligible with respect to the bearing length $L$).

The domain length and width are $L_{dom}=8.48\times10^{-4}$m and $W_{dom}=4.2\times10^{-5}$m, respectively. The slider length is $L=4.24\times10^{-4}$m. The initial position of the plate barycentre is represented by the horizontal coordinates $x_{CM,0}=0.35\times L_{dom}=2.968\times10^{-4}$m, $y_{CM,0}=W_{dom}/2=2.1\times10^{-5}$m. The dynamic viscosity is $\mu=319\times10^{-3}$Pa×s, representative of the motor oil "SAE 40" at ambient conditions; the oil density is $\rho=900$kg/m$^3$, the oil depth is $h=0.021$mm. The upstream and downstream fluid frontiers are represented by open boundaries. Monitors are located along the domain centreline ($y=W_{dom}/2$).

The spatial resolution is defined by $dx=2.1\times10^{-6}$m, $h_{SPH}/dx=1.3$ and $dx/dx_s=2$. Since the spatial discretization of the numerical pressure profile is uniform, the numerical non-dimensional load bearing capacity is estimated as the average of the pressure coefficient values over the mobile plate bottom:

$$\langle L_C \rangle \equiv \frac{\langle l_c \rangle}{\frac{1}{2}\rho(U^*)^2 L} = \frac{2}{\rho(U^*)^2 L}\sum_{i=1,N} p_{s,i}dx_i = \frac{1}{N}\sum_{i=1,N} C_{p,s,i} \qquad (62)$$

where $N$ represents the number of surface body particles along the centreline of the mobile plate bottom.

The stationary regime for this uniform slider is obtained at ca. $t_f=2.4\times10^{-6}$ s and a laminar regime is detected in the fluid volume within the bearing. At this time, turbulence only affects the leading and the trailing edge regions. For further times, turbulence might progressively interest a bigger fluid volume within the plates due to the propagation of the fluid-structure interactions at the plate edges.

Results are reported in terms of non-dimensional quantities. Figure 2 (top panel) shows a lateral view of the 3D field of the x-component of the normalized velocity. No-slip conditions are well reproduced both along the fixed frontier and along the plate bottom. The homogeneity along the *x*-axis is perturbed around both the leading and the trailing edges of the moving plate. Figure 22 (centre panel) shows a lateral view of the 3D field of the pressure coefficient. One notices that the presence of a leading and a trailing edge is not coherent with imposing null pressure values at the inlet and outlet section of a linear slider under stationary regime: the leading/trailing edge is locally featured by an over/under-pressure region. In particular, in the presence of a free surface flow, the liquid lifts up at the leading edge and the upstream liquid is slower than the fluid film below the plate. The simulation shows a local turbulent regime at the plate edges and highlights a drawback of Reynolds' equation for fluid films at both the inlet and outlet sections of the slider. Figure 22 (bottom panel) shows a 3D view of the field of the *x*-component of the normalized velocity. Although the configuration of this uniform slider is 2D, the model is able to represent an equivalent 3D configuration, with a proper representation of the fluid volume, the solid volume, the fixed bottom and the symmetry planes. This feature is relevant for its applicability to complex topological configurations, see Sec. 8.

Figure 3 (left and centre panel) reports the code validations for the uniform slider on the pressure coefficient longitudinal profile. Four numerical plots are reported: the reference one is monitored along the surface body particles representing the fluid-body interface. The other plots are monitored within the fluid domain at different non-dimensional heights ($Z=z/h$): $Z_1=1$, $Z_2=0.5$, $Z_3=0$. Results are plotted far from the slider edges so that the effective slider length is $L_{eff}=0.95L$. The SPH code can closely reproduce the linear $C_p$ longitudinal profile of the analytical solution (Sec. 3). Close to the plate leading edge, where some minor differences appear, the hypotheses of Reynolds' equations for fluid films are not fully respected, as also observed by Vakilian et al. [3] and Dobrica and Fillon [2]. Contrarily to the analytical solution, the code can simulate the *z*-dependent pressure field, the velocity vertical component and the inertial effects due to the fluid-structure interactions at the leading and the trailing edges. Further, the inter-comparisons between the SPH profiles at different heights show the accuracy of the code in reproducing the uniformity of the pressure field along *z*-axis, far enough from the plate edges (Fig. 3, centre panel).

Figure 3 (right panel) reports the code validations for the uniform slider on the vertical profiles of the *x*-component of the normalized velocity. The numerical probes are located at $X_1$=0.26, $X_2$=0.5 and $X_3$=0.74, where $X$ is the non-dimensional distance from the trailing edge $(x-x_0)/L$. The agreement between the numerical model and the analytical solution is very good. The boundary treatment related to moving solid bodies (Sec. 4.3, domain top) seems more accurate than the method related to fixed boundaries (Sec. 4.1, domain bottom).

Table 1 (second row) reports the estimation ($L_{C,SPH}$=0.445) for the non-dimensional load capacity and its analytical solution ($L_{C,an}$=0.452). The SPH relative error on this quantity is 1.74%.

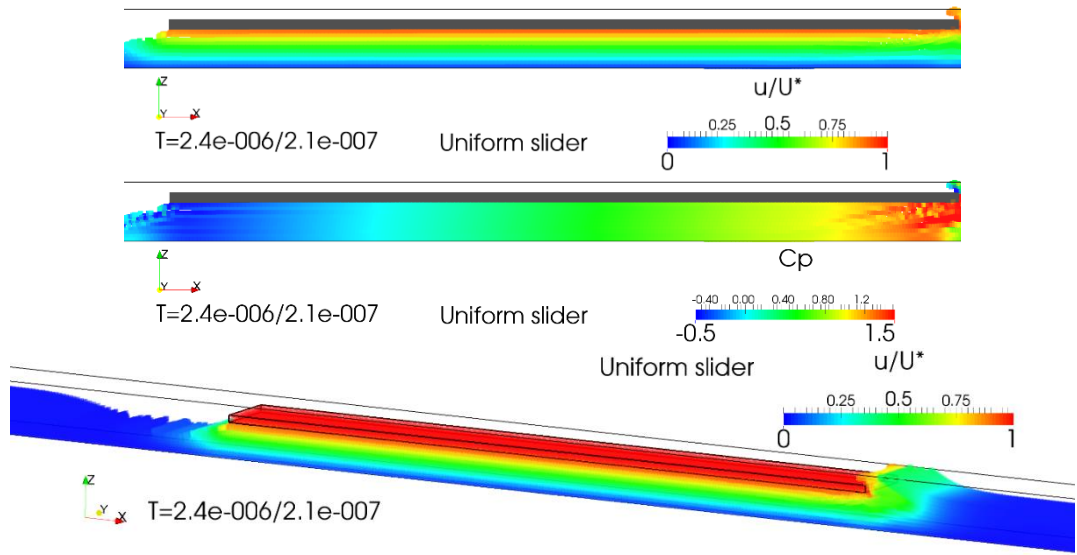

Figure 2. Uniform slider. Top panel: 3D field (lateral view) of the *x*-component of the normalized velocity. Centre panel: 3D field (lateral view) of the pressure coefficient (at the leading and trailing edges the values go beyond the legend estrems). Bottom panel: 3D field (3D view) of the *x*-component of the normalized velocity.

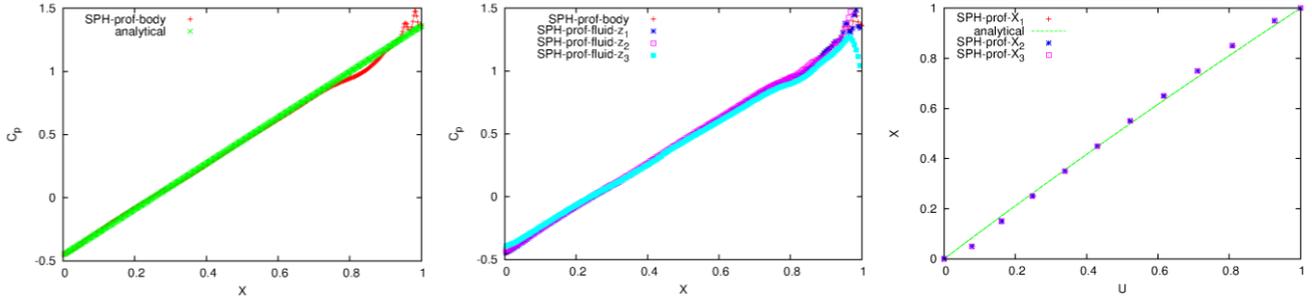

Figure 3. Validations for the uniform slider: pressure coefficient longitudinal profile (comparison with measures -left panel-; proof of vertical homogeneity -centre-) and vertical profiles of the *x*-component of the normalized velocity.

| Bearing type | $L_C$ (SPH) | $L_C$ (analytical) | Relative error (%) |
|---|---|---|---|
| Uniform slider | 0.445 | 0.452 | 1.74 |
| Linear slider | 2.77 | 2.69 | 2.97 |

Table1. Non-dimensional load capacity: SPH estimations, analytical solutions and relative errors for the uniform slider and the linear slider.

## 7 Validation: linear slider

Figure 4 shows an example of the velocity field for the linear slider under dynamic stationary conditions for both the fluid and the solid sub-domains. The fluid depth shows a uniform rate of change for the water depth along *x*-axis. The velocity gradient grows with the distance from the leading edge.

With respect to the uniform slider discussed in Sec. 6, the following input data have to be modified to simulate a linear slider. The maximum and minimum values of the fluid depth are $h_{max}=2.10\times10^{-5}$m and $h_0=1.68\times10^{-5}$m, respectively ($k=h_{max}/h_0-1=0.25$, with a slope angle of 0.568°, which is relevant as $h<<L$). The height of the plate barycentre is lowered of the quantity $kh_0/2=2.1\times10^{-6}$m during the rotation of the plate around the bottom side of the leading edge: this occurs during the first 5% of the simulated time, while the plate is translating along *x*-axis. Stationary conditions are dynamically achieved at $t_f=6.0\times10^{-6}$s ($T_f=17.9$).

The following features (see Fig. 4 where an example of the velocity field for the linear slider under stationary conditions achieved at the end of the dynamic simulations is shown, for both the fluid and the solid sub-domains) are analogous to the uniform slider: no-slip conditions are well reproduced at both the fixed frontier and the plate bottom; the homogeneity of the pressure field along the *x*-axis is perturbed only close to the leading and the trailing edges of the mobile plate; the model is able to represent an equivalent 3D configuration of the 2D analytical slider.

Figure5 (left panel) reports the code validation on the linear slider bearing in terms of pressure coefficient ($C_p$) longitudinal profile. Profiles are plotted far from the slider edges so that the effective slider length is $L_{eff}=0.95L$. The SPH profile fairly matches the herein analytically derived solution. The maximum value of the pressure coefficient is higher than unity because an external force continuously applies to the fluid domain. Figure (right panel) demonstrates a good representation of the vertical homogeneity of the pressure field, by means of comparison between the probes within the fluid domain at the different non-dimensional heights $Z_1=1$, $Z_2=0.5$ and $Z_3=0$.

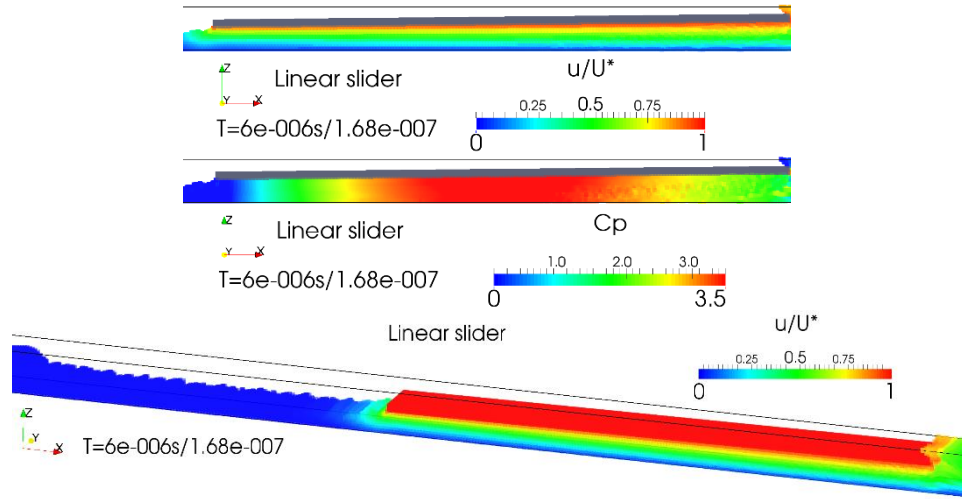

Figure 4. Linear slider. 3D fields. Top panel: non-dimensional velocity (lateral view). Centre panel: pressure coefficient (lateral view). Bottom panel: non-dimensional velocity (3D view).

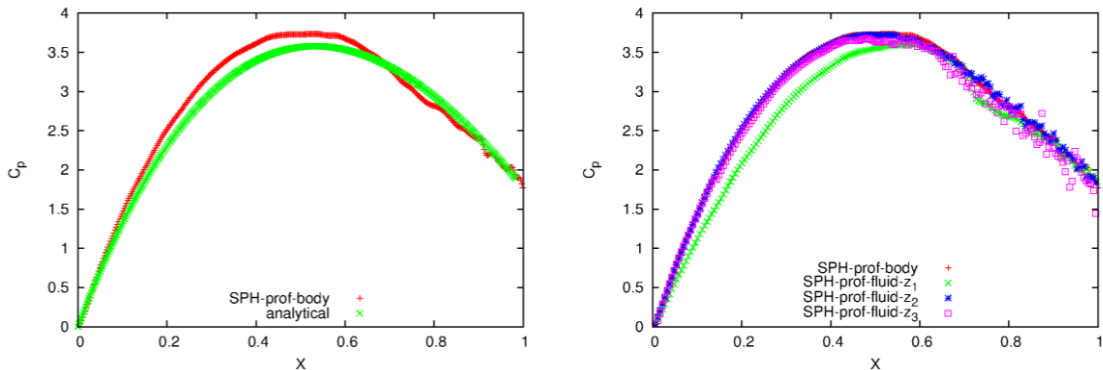

Figure 5. Validations for the linear slider: pressure coefficient longitudinal profile. Comparison with measures (left panel) and proof of vertical homogeneity (right panel).

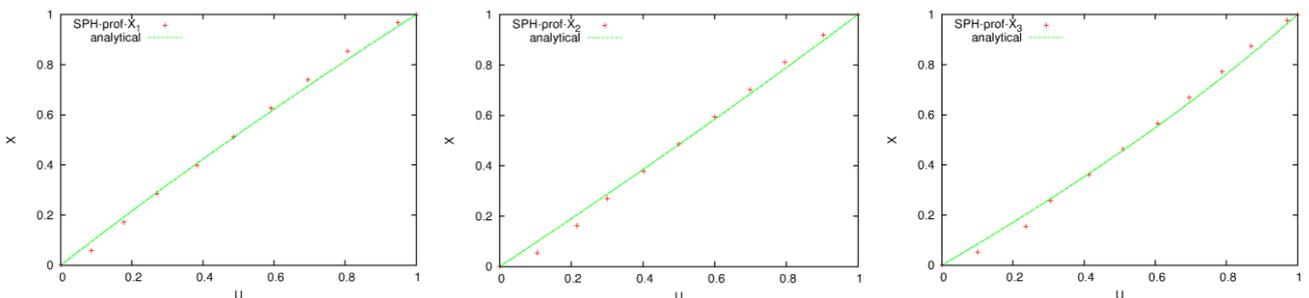

Figure 6. Validations for the linear slider: vertical profiles of the *x*-component of the non-dimensional velocity.

Figure 66 shows the code validations on the linear slider bearing in terms of velocity vertical profiles. The numerical SPH profiles fairly agree with the non-linear analytical solutions at the three monitoring probes $X_1=0.39$, $X_2=0.64$ and $X_3=0.88$. Contrarily to the uniform slider, the vertical profile of the horizontal velocity is quadratic in $z$ and depends on $x$. The SPH code is able to represent an upward concavity (towards the leading edge, $X=X_3$, right panel) and a downward concavity towards the trailing edge ($X=X_1$, left panel). Analogously to the uniform slider, the boundary treatment for the solid bodies (top region of the profiles) seems slightly more accurate than the semi-analytical approach (bottom region of the profiles).

Table 1 (second row) reports the SPH estimation ($L_{C,SPH}=2.77$) for the non-dimensional load capacity and its analytical solution ($L_{C,an}=2.69$), with a relative error of the code equal to 2.97%.

**8 Applicability of the computational framework to complex rough surfaces**

The linear slider bearing of Sec. 7 is herein modified by imposing a complex 3D surface as the bottom fixed frontier, to demonstrate the versatility of the approach to take into account complex rough surfaces as input data. Its shape is taken from a real natural geometry of a strawberry leaf that was acquired using the Leica DCM 3D confocal profilometer in the MUSAM-Lab of the IMT School for Advanced Studies Lucca. The spatial resolution needs be finer than the previous test cases due to roughness, with $dx=1.66\times10^{-6}$m.

Figure 7 shows an example of the velocity field and highlights the capabilities of the code in simulating complex fluid-structure interactions in the field of hydrodynamic lubrication. Further investigation on the effect of roughness on the velocity and pressure fields is left for further investigation in a dedicated follow-up article.

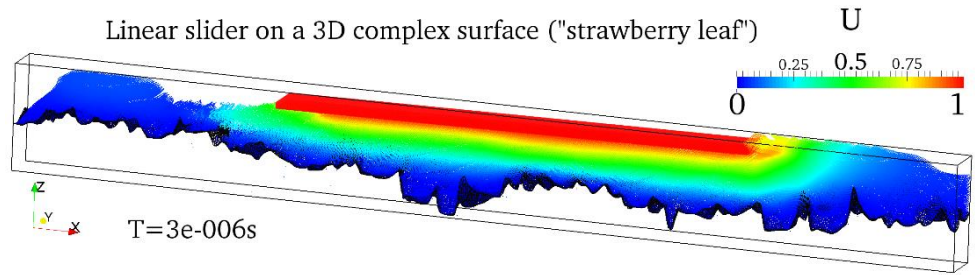

Figure 7. Linear slider on a 3D complex surface. Example of a 3D field of the non-dimensional velocity (3D view).

**9 Conclusions**

SPHERA has been herein empowered to deal with 3D "fluid - solid mobile body" interactions under no-slip conditions and laminar regimes. The code has been validated in relation to analytical solutions for uniform and linear slider bearings, herein generalized for any Dirichlet's boundary condition. With respect to the state-of-the-art (i.e., 2D codes based on Reynolds' simplified equation for fluid films), SPHERA showed the following features and competitive advantages: (i) 3D complete formulation of the Navier-Stokes equations for incompressible fluids with uniform viscosity, without the need of making simplified assumptions on the fluid flow as in Reynolds' simplified models for fluid films; (ii) validations on both local and global quantities (pressure and velocity profiles; load capacity); (iii) possibility to simulate hydrodynamic lubrication in the presence of complex 3D rough topologies, where semi-analytical solutions are not available. This study also showed the advantages of using a CFD-SPH code in simulating the inertia and 3D effects close to the slider edges. All of these features do not belong to the codes for hydrodynamic lubrication based on Reynolds' equation for fluid films.

This study allowed SPHERA to deal with hydrodynamic lubrication and empowered the code on other application fields with fluid-structure interactions (e.g., transport of solid bodies by floods and earth landslides; rock landslides). SPHERA is developed and distributed on a GitHub public repository.

**Acknowledgements**

We acknowledge the CINECA award under the ISCRA initiative, for the availability of High Performance Computing resources and support." In fact, SPHERA simulations have been financed by means of the following instrumental funding HPC project: HSPHCS9; HSPHER9b.

The contribution of the author of RSE SpA has been financed by the Research Fund for the Italian Electrical System (for "Ricerca di Sistema -RdS-"), in compliance with the Decree of Minister of Economic Development April 16, 2018.

SPHERA v.9.0.0 is realised by RSE SpA thanks to the funding "Fondo di Ricerca per il Sistema Elettrico" within the frame of a Program Agreement between RSE SpA and the Italian Ministry of Economic Development (Ministero dello Sviluppo Economico).